# Implementation of Security in Distributed Systems – A Comparative Study

Mohamed Firdhous
Faculty of Information Technology,
University of Moratuwa,
Moratuwa,
Sri Lanka.
Mohamed.Firdhous@uom.lk

*Abstract* – **This paper presents a comparative study of distributed systems and the security issues associated with those systems. Four commonly used distributed systems were considered for detailed analysis in terms of technologies involved, security issues faced by them and solution proposed to circumvent those issues. Finally the security issues and the solutions were summarized and compared with each other.**

I. INTRODUCTION

In today's networked world, computers rarely work in isolation. They collaborate with each other for the purpose of communication, processing, data transfer, storage etc., When systems work in this collaborative fashion with other systems that are geographically scattered over wide distance it is commonly known as a distributed system. In literature, researchers have used diverse definitions to outline what a distributed system is.

Coulouris et al., have defined a distributed system as "a system where the hardware and software components have been installed in geographically dispersed computers that coordinate and collaborate their actions by passing messages between them [1]. Tanenbaum and Van Steen have defined a distributed system as "a collection of systems that appears to the users as a single system" [2]. From Tanenbaum's definition, it can be conceived that a distributed system refers to a software system rather than the hardware that are involved in creating the system. Combining these definitions, it can be stated that a distributed system is an application that communicates with multiple dispersed hardware and software in order to coordinate the actions of multiple processes running on different autonomous computers over a communication network, so that all components hardware and software cooperate together to perform a set of related tasks targeted towards a common objective.

Most people consider a distributed system and a network of computers to be the same. But these two terms mean two different but related things. A computer network is an interconnected set of autonomous computers that communicated with each other. A user using a computer network understands that he uses different resources lying on different computers as a computer network does not hide the existence of multiple computers. But a distributed system on the other hand provides the feeling that the user is working on a single homogenous more powerful computer with more resources. The existence of multiple autonomous computers is transparent to the user as the distributed system application that is running on the computers would select suitable computers and allocate jobs without the specific intervention of the user [3].

Distributed systems have been built with the objective of attaining the following:

- Transparency
- Openness
- Reliability
- Performance
- Scalability

In order to achieve the above objectives, security of the system must be given adequate attention as it is one of the fundamental issues in distributed systems [4]. Attention must be paid at every stage including design, implementation, operation and management of distributed systems.

In this paper, the author takes an in depth look at the implementation of security in some most popular distributed systems.

II. DISTRIBUTED SYSTEMS

There are many distributed systems in operation today. The following are some of the most popular distributed systems in use today.

- Cluster Computing
- Grid Computing
- Distributed storage systems
- Distributed databases

*A. Cluster Computing*

Computers communicating over a high speed network can be made to work and present itself as a single computer to the users. A set of computers that are grouped together in such a manner that they form a single resource pool is called a cluster. Any task that has been assigned to the cluster would run on all the computers in the cluster in a parallel fashion by breaking the whole task into smaller self contained tasks. Then, the result of the smaller tasks would be combined to form the final result [5].

Cluster computing helps organizations to increase their computing power using the standard and commonly available technology. These hardware and software which are commonly known as commodity items can be purchased from the market at relatively low cost [6]. Cluster computing has seen tremendous growth in the recent years. Around 80 percent of top 500 supercomputing centers in the world are using clusters. Clusters are used primarily to run scientific, engineering, commercial, and industrial applications that require high availability and high throughput processing [7]. Protein sequencing in biomedical applications, earth quake simulation in civil engineering, petroleum reservoir simulation in earth resource and petroleum engineering and replicated and distributed storage and backup servers for high demand web based business applications are a few examples for applications which primarily run on clusters [8-11]. Figure 1 shows a typical arrangement of computers in a Computing Cluster.

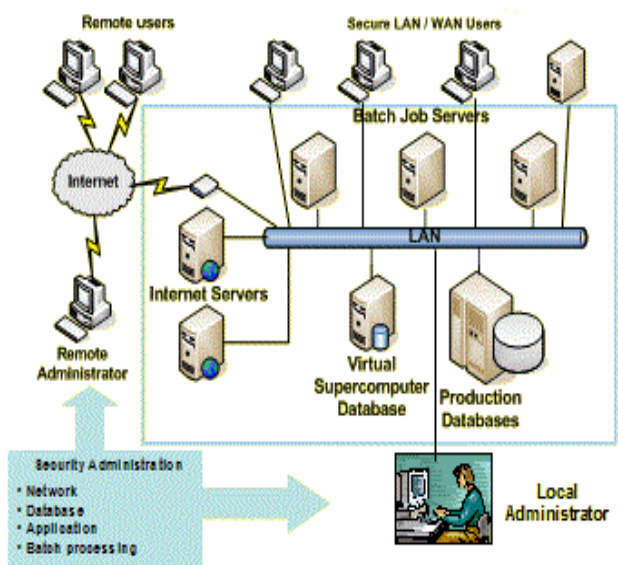

Figure 1: Computing Cluster

*B. Grid Computing*

Grid is a type of distributed computing system where a large number of small loosely coupled computers are brought together to form a large virtual supercomputer. This virtual super computer has to perform tasks that are large for any single computer to perform within a reasonable time.

Grid is defined as a parallel and distributed system that is capable of selecting, sharing, and aggregating geographically distributed resources dynamically at runtime based on their availability, capability, performance, and cost meeting the users' Quality of Service (QoS) requirements [12]. Grid computing combines computing resources distributed across a large geographical area belonging to different persons and organization. The main purpose of the grid system is to collaboratively work across multiple systems to solve single computing task by dividing the task into smaller self contained tasks and distributing those tasks to different computers.

The middleware used in grid computing is responsible for dividing and apportioning the tasks. The size of a grid system can vary from few hundred computers within an organization to large systems consisting of thousands of nodes across multiple organizations. Small grids confined to a single organization is commonly known as intra-node corporation while the larger wider system is referred to as inter node corporation [13]. Figure 2 shows Grid System distributed across heterogeneous computing platforms.

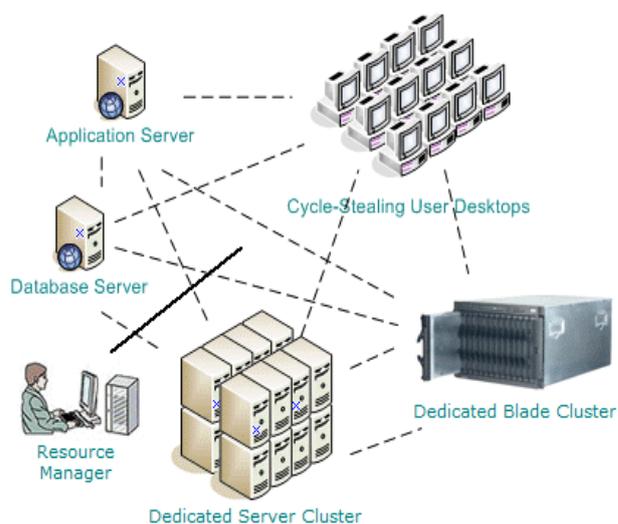

Figure 2: Grid Computing System

Grids have been used to perform computationally intensive scientific, mathematical, and academic problems through volunteer computing. Drug discovery, economic forecasting, seismic analysis, and back office data processing for e-commerce are a few of the tasks that are commonly solved using grid computing.

*C. Distributed Storage Systems*

The rapid growth of storage volume, bandwidth and computation resources along with the reduction in the cost of storage devices have fueled popularity of distributed storage systems. The main objective of distributing storage across multiple devices is to protect the data in case of disk failure through redundant storage in multiple devices and to make data available closer to the user in massively distributed system [14]. There are mainly four types of distributed storage systems. There are namely, Server Attached Redundant Array of Independent Disks (RAID), centralized RAID, Network Attached Storage (NAS) and Storage Area Network (SAN) [15]. NAS and SAN are the most popular distributed storage techniques out of the four.

Figure 3 shows the typical arrangement of distributed storage system.

## III. SECURITY IN DISTRIBUTED SYSTEMS

Security is one of the most important issues in distributed systems. When data is distributed across multiple networks or information is transferred via public networks, it becomes vulnerable to attacks by mischievous elements. Similarly other computing resources like processors, storage devices, networks etc., can also be attacked by hackers.

### A. Security for Computing Clusters

When the computing clusters are made available to the public or networks are setup using public resources such as the Internet, they become subject to various kinds of attacks. The most common types of attacks on the clusters are computation-cycle stealing, inter-node communication snooping, and cluster service disruption [17]. Hence the clusters have been protected by security mechanisms that include services like authentication, integrity check, and confidentiality. The main purpose of the security mechanisms is to protect the system against hackers as well as to meet the security requirements of the applications.

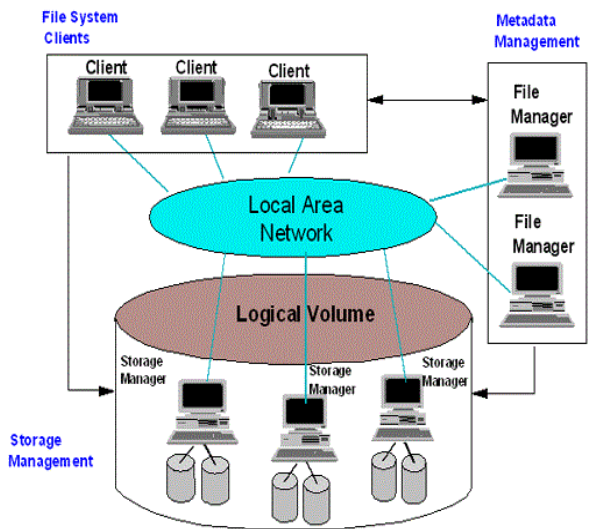

Figure 3: Distributed Storage System

NAS and SAN have slight differences in techniques adopted for transferring data between devices and the performance due to this difference. NAS mainly uses TCP/IP protocol to transfer data across multiple devices whereas SAN uses SCSI setup on fiber channels. Hence NAS can be implemented on any physical network supporting TCP/IP such as Ethernet, FDDI, or ATM. But SAN can be implemented only fiber channel. SAN has better performance compared NAS as TCP has higher overhead and SCSI faster than TCP/IP networks.

Li and Vaughn have studied the security vulnerabilities of computing clusters using exploitation graphs (e-graphs). They have modeled several attacks that can be carried on all three pillars of security namely, confidentiality, integrity and availability. They have shown that e-graphs can be simplified based on domain knowledge such as cluster configurations, detected vulnerabilities, etc. they further state that this technique could be used for certification of clusters with the help of a knowledge base of cluster vulnerabilities[18].

Xie and Qin have developed two resource allocation schemes named Deadline and Security constraints (TAPADS) and Security-Aware and Heterogeneity-Aware Resource allocation for Parallel jobs (SHARP). These two schemes ensure that parallel applications executed on computing clusters meet the security requirements while meeting the deadline of executions [17]. Hence it could be seen that if these schemes ensure mainly the availability of the system as timely execution of an application is an indication of the availability of the resources.

### D. Distributed Database System

Distributed database system is a collection of independent database systems distributed across multiple computers that collaboratively store data in such a manner that a user can access data from anywhere as if it has been stored locally irrespective of where the data is actually stored [16]. Figure 4 shows an arrangement of distributed database system across multiple network sites.

Denial of Service (DoS) attack is one of the common attacks on distributed systems. These attack mainly target resources in such a manner that the resources are prevented from carrying out their legitimate operations. A method that uses services and markov chain to mitigate the effects on the DoS attack on a cluster based wireless sensor network has been presented in [19].

Hence it can be seen that computing clusters are vulnerable to attacks by mischievous elements like hackers and crackers due to its open nature and use of public resources such as the internet. Extensive research has been carried out by several researchers on the security of clusters and they have proposed several methods that can be made used to protect the clusters from these attacks.

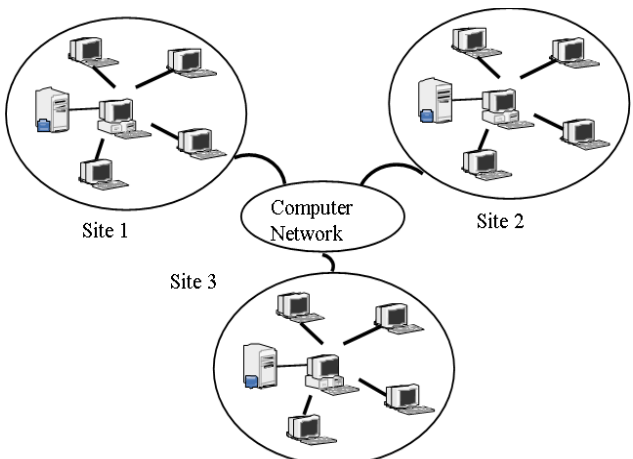

Figure 4: Distributed Database System

## B. Grid System Security

Grid computer systems provide several security mechanisms to protect the grid resources against attacks. Middleware is one of the critical system software in the grid infrastructure as it provides the common communication infrastructure and makes the grid services available to applications. Middleware also allows for a uniform security configuration at the service container or messaging level. Grid authentication is based on Public Key Infrastructure (PKI) and capable of handling different types of user credentials such as PKI, SAML, Kerberos tickets, password, etc., Delegation is one of the necessary mechanisms in grid service delivery and is implemented using X.509 Proxy Certificate. Authorization to access grid resources is based on Virtual Organization (VO) attributes assigned to a user and managed by Virtual Organization Membership Service (VOMS). Trust management in grid systems are handled using certificates and trust relations are represented by a certificate chain that include Grid Certification Authority (CA) certificate and other successively generated proxies [20].

Grid authentication module is one of the critical components in preventing external users from randomly accessing internal grid and protecting the grid system from unauthorized users. This module handles security threats from internal network, when certificated grid users carry out illegal (unauthorized) operations within the grid [21].

These grid security mechanisms are all implemented on almost all grid systems available today. There several grid community initiatives going on in the area of grid middleware interoperability which would finally unify the grid security as a single coherent security platform and scheme.

## C. Distributed Storage System Security

Several active researches are going on in the area of threat modeling and developing security model for protecting distributed storage systems. The most important resource in the distributed storage system is the data stored in the storage devices of the system. This data needs to be properly labeled and protected. Also any protection system introduced must be backward compatible in other words; it not only should protect the data stored after the security scheme is installed but also the data that had been there prior to the introduction of that scheme.

Hasan et al., have introduced a threat model named CIAA threat model. This model addresses all the security issues namely, Confidentiality, Integrity, Availability and Authentication. In arriving at this model, authors have organized the threats on a distributed storage system under each category of the CIAA pillars of security and provided techniques that can be used to circumvent the threats. The other security model discussed by the authors is the Data Lifecycle Model that examines the types of threats that may occur at different stages of data state from creation to extinction. Under this model threats have been organized under six groups and solutions have been proposed [22].

Dikaliotis, Dimakis and Ho have proposed a simple linear hashing technique that can detect errors in the storage nodes in the encoded distributed storage systems [23]. Mutually Cooperative Recovery (MCR) mechanism enables the system to recover data in situations of multiple node failures. The transmission scheme and design a linear network coding scheme based on (n, k) strong-MDS code proposed help recover systems from failure with relative ease [14].

Hence it can be seen that the security schemes in the distributed storage systems mainly concentrate on data security in terms of integrity and failure management (availability).

## D. Distributed Database Security

Distributed database management systems face more security threats compared to their counterpart centralized database systems. The development of security for distributed database systems have become more complicated with the introduction of several new database models such as object-oriented database model, temporal database model, object relational database model etc.

In traditional security model, all the data stored in database and the users who access that data belong to the same security level. A multilevel secure database system assigns security level to each transaction and data. Clearance level of a transaction is represented by security level assigned to it and the classification level of data is given by the classification level. A multilevel secure database management system (MLS/DBMS) restricts database operations based on the security levels [24]. From the above discussion, it can be seen that by introducing the military information classification and access control security of distributed databases can be enhanced.

Zubi has presented a design that would improve the scalability, accessibility and flexibility while accessing various types of data in a distributed database system. He has also proposed multi level access control, confidentiality, reliability, integrity and recovery to manage the security of a distributed database system [25].

## IV. SUMMARY

From the above discussion, it can be seen that security becomes more prominent when the systems have been distributed across over multiple geographic locations. Each type of distributed system has its own peculiar security requirements. But, all the systems have the common CIA triad as the heart of any security implementation. In computing clusters and grids the security mainly concentrates on protecting the data in transit and access to distributed resources. Security in clusters is somewhat simpler compared to grid due to homogeneous nature of clusters. One of the main attacks that has been carried out on clusters is the Denial of Service (DoS) attack. Researchers have proposed novel methods based on markov chain to mitigate the impact of DoS attacks.

In grid the middleware layer provides the platform for the implementation of security on the entire grid system. Grid system use strong security based on PKI and X.509 certificates. The user authentication module in the grid provides security against threats by external sources and illegal actions by internal users.

Security of distributed storage systems mainly concentrate on securing data. The main areas concentrated on distributed storage are protection against data corruption and protection of data in situations of node failures. Researchers have proposed various models and schemes to protect the storage system against attacks and node failures.

In distributed database system, the security implementation has been made more complicated due to the availability of different kinds of database models. But researchers have shown that by applying multi level security based on military information classification and access control, distributed database security can be enhanced.

## V. CONCLUSION

In this paper, the development of distributed systems was discussed in terms of what a distributed system is and the objectives of setting up a distributed system. From all the available distributed systems, four most commonly used distributed systems were discussed in depth and then the security issues faced by these systems and the solutions proposed by various researchers were discussed in depth. Finally the security issues and solutions proposed for different systems were summarized and compared with each other.

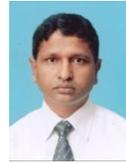
**Mohamed Fazil Mohamed Firdhous** is a senior lecturer attached to the Faculty of Information Technology of the University of Moratuwa, Sri Lanka. He received his BSc Eng., MSc and MBA degrees from the University of Moratuwa, Sri Lanka, Nanyang Technological University, Singapore and University of Colombo Sri Lanka respectively. In addition to his academic qualifications, he is a Chartered Engineer and a Corporate Member of the Institution of Engineers, Sri Lanka, the Institution of Engineering and Technology, United Kingdom and the International Association of Engineers. Mohamed Firdhous has several years of industry, academic and research experience in Sri Lanka, Singapore and the United States of America.
`